\documentclass{article}
\usepackage[utf8]{inputenc} 
\usepackage{arxiv}
\usepackage{graphicx}
\usepackage{amsmath}
\usepackage{algorithm} 
\usepackage{algpseudocode}
\usepackage{plex-serif}
\usepackage[basic,italic,symbolgreek]{mathastext}
\usepackage{hyperref}
\hypersetup{colorlinks,allcolors=black}
\begin{document}

\title{NINEPINS: Nuclei Instance Segmentation with Point Annotations}

\author{
Ting-An Yen\\
National Cheng Kung University
\And
Hung-Chun Hsu\\
National Cheng Kung University \And
Pushpak Pati\\
IBM Research Europe \And
Maria Gabrani\\
IBM Research Europe \And
Antonio Foncubierta-Rodríguez\\
IBM Research Europe \And
Pau-Choo Chung\\
National Cheng Kung University
}

\maketitle              
\begin{abstract}
Deep learning-based methods are gaining traction in digital pathology, with an increasing number of publications and challenges that aim at easing the work of systematically and exhaustively analyzing tissue slides. 
These methods often achieve very high accuracies, at the cost of requiring large annotated datasets to train. 
This requirement is especially difficult to fulfill in the medical field, where expert knowledge is essential.
In this paper we focus on nuclei segmentation, which generally requires experienced pathologists to annotate the nuclear areas in gigapixel histological images. 
We propose an algorithm for instance segmentation that uses pseudo-label segmentations generated automatically from point annotations, as a method to reduce the burden for pathologists.
With the generated segmentation masks, the proposed method trains a modified version of HoVer-Net model to achieve instance segmentation. 
Experimental results show that the proposed method is robust to inaccuracies in point annotations and comparison with Hover-Net trained with fully annotated instance masks shows that a degradation in segmentation performance does not always imply a degradation in higher order tasks such as tissue classification.

\end{abstract}
\section{Introduction}
Deep learning has shown its potential in achieving accuracy levels beyond the state of the art of classical computer vision methods in tasks like image classification, object detection, object segmentation and so on.
For example, in~\cite{AlexNet,RCNN,YOLO,FCN}, deep learning models outperform traditional methods in the experiments.
Deep learning is also widely used for medical image processing tasks, e.g. cell and membrane segmentation~\cite{UNet}, breast cancer mitosis detection~\cite{AggNet}, and so forth.
However, supervised training of deep learning models requires a large amount of annotations. 
This can be specially problematic in domains where annotation requires expert knowledge, like in medical imaging. 

In digital pathology, nuclei segmentation is a common task for many diseases, and many high order tasks such as tumor grading or classification rely on accurate nuclei segmentation in order to describe the tissue~\cite{cgc}. 
However, the labeling of nuclei segmentation masks needs to be done by experienced pathologists because the appearance of different types of cells differs largely and the boundaries of nuclei are difficult to identify, which renders the annotation very time and economically expensive.  

In this paper we explore the issue of minimizing the annotation burden required for nuclei segmentation. 
We propose a modification of the state of the art on nuclei segmentation, HoVer-Net~\cite{hover}, called NINEPINS (Nuclei INstance sEgmentation from PoInt aNnotationS) that generates pseudo-label segmentation masks for training. 
We analyze the trade-offs incorporating a varying proportion of pseudo-labels with respect to true labels during training. 
Additionally, we test the robustness and the tolerance of point inaccuracies.
Finally, to evaluate the impact of utilizing this semi-supervised framework in higher order tasks, this paper uses CGC-Net ~\cite{cgc}, which was proposed by Zhou \textit{et al.}, as a classification model using the nuclear instance segmentation results as inputs to classify tumor tissue from the Camelyon16~\cite{camelyon} dataset.


\section{Related work}
\label{sec:related_work}
Since annotation availability is a very important limiting factor for the adoption of deep learning based methods in digital pathology, other researchers have already addressed this issue. 
To reduce the annotation cost, Hui \textit{et al.}~\cite{weak} proposed a weakly-supervised method to train a nuclei segmentation model with point annotations instead of mask annotations. 
Point annotations are used to represent the approximate centers of nuclei. 
Since only one click for each nucleus is needed and the boundaries don't need to be identified, point annotations enormously reduce the effort required to annotate nuclei.
The paper proposed an approach to generate mask annotations from point annotations using the traditional computer vision techniques with prior knowledge of the target task. Additionally, a dense conditional random field (CRF) loss was proposed to refine the segmentation results. 
PseudoEdgeNet~\cite{pseudo}, which was proposed by Yoo \textit{et al.}, is another example of weakly-supervised training. Instead of generating mask annotations, the work trained the segmentation model directly with point annotations. 
To make the segmentation prediction grow from point to nuclear area, an additional sub-network called PseudoEdgeNet was attached to the model to predict nuclear edges. 
The predicted edges were further used in the computation of edge loss which compared them with the edges of segmentation result. 
Nevertheless, these methods only produced binary segmentation results which are not sufficient when features of individual nucleus are required but nuclei are overlapping each other. In such cases, an instance segmentation algorithm is needed.

For instance segmentation of nuclei, Simon \textit{et al.}~\cite{hover} proposed a model called HoVer-Net which predicted segmentation mask and distance maps at once. Furthermore, the results were combined together to produce instance segmentation mask. In the paper, classification of nuclei can be part of the framework, which is an example when instance segmentation is needed rather than only binary segmentation. 
It is straightforward that by integrating~\cite{weak} and~\cite{hover} one can achieve instance segmentation with only point annotations. 
However, to train the distance map branch in HoVer-Net, distance map labels were required, which were generated from instance mask annotations. 
In the case that only point annotations are accessible, there needs to be another way to produce distance map labels.

Our method aims to segment individual nuclei from digital pathology images. 
An algorithm is proposed to improve the segmentation labels generated by~\cite{weak} and to produce the distance map labels. 
Additionally, this paper uses a modified version of HoVer-Net as the instance segmentation model. 
Thus, as a whole, this paper proposes a method as a semi-supervised learning framework for training an instance segmentation model with only point annotations.


\section{Methods}
\subsection{Instance Segmentation Network}
The architecture of models in this paper is based on HoVer-Net in \cite{hover}. 
It is composed of one encoder and optional amount of decoder branches. 
The encoder is the same as the one of ResNet-50~\cite{ResNet} and loads weights pretrained with ImageNet~\cite{ImageNet} when training, and the decoders are the identical to each other but for different purposes. 
For the HoVer-Net trained in this paper, two decoder branches for segmentation and prediction of distance maps are used, but the classification decoder branch in \cite{hover} is disabled in the following experiments. 
Additionally, another proposed model contains a third decoder branch for nuclear centroid detection. 
The ground truth for this branch is point labels representing nuclear centroids. 

\subsection{Pseudo-label}
In~\cite{weak}, Voronoi diagram and K-means clustering were used to generate distance-based labels and color-based labels respectively. The author trained the model with both types of labels at the same time to balance the weak points of the them, i.e. distance-based labels didn't describe the area of nuclei and color-based labels usually overestimated the area of nuclei and also had a large number of false positives. Although each of the two types of labels was very different from the true labels alone, they could yield a better approximation to the true labels when combined together properly.

To make pseudo-labels better approximation to true labels, connected components not covering any dilated point were removed and points not covered by pseudo-labels were added to the nuclei masks after dilated.

\begin{algorithm}
	\caption{Pseudo-label refinement} 
	\begin{algorithmic}[1]
		\For {connected component $CC$ in pseudo-label $Ps$}
			\If{$\left | CC\cap dilate(P, 5)) \right | = 0$}
			\State $Ps \leftarrow Ps - CC$
			\EndIf
		\EndFor
		\For {point $p$ in point label $P$}
			\If{$\left | p\cap Ps \right | = 0$}
			\State $Ps \leftarrow Ps \cup dilate(p, 3)$
			\EndIf
		\EndFor
	\end{algorithmic} 
where $dilate(A, b)$ represents an operation which dilates objects in $A$ with a $b$-radius circle as the structure element and $\left | M\right |$ calculates the area of $M$.
\end{algorithm}

\subsection{Distance Map Labels}
To train the HoVer branch, distance map labels are required. In~\cite{hover}, such labels are generated from instance labels. However, in the targeted case, only point labels and pseudo-labels (for segmentation) are available. In this paper, Voronoi edges in Voronoi diagram are used to separate segments in pseudo-labels and to generate pseudo-instance-labels.

\section{Experiments and results}

\subsection{Metrics}
In the following experiments, performance is evaluated in terms of two metrics, DICE and $\mathrm{DQ}_{point}$. DICE is a metric proposed in~\cite{DICE} which is commonly used to estimate the segmentation quality and has the formula:

\begin{equation}
DICE = \frac{2 * \left | X \cap Y \right |}{\left | X \right | + \left | Y \right |}.
\end{equation}

where $X$ and $Y$ represent model prediction and ground truth label respectively. $\mathrm{DQ}_{point}$ is modified from Detection Quality (DQ), which is proposed in~\cite{hover} and has the following form:

\begin{equation}
DQ = \frac{\left | TP \right |}{\left | TP \right | + \frac{1}{2}\left | FP \right | + \frac{1}{2}\left | FN \right |}.
\end{equation}

where $TP$ stands for ground truth centroids which are covered by a unique prediction segment (one prediction segment can only be assigned to one centroid), instead of paired segments with $IOU > 0.5$ in~\cite{hover}.

\subsection{Datasets}
For evaluation and comparison, colorectal nuclear segmentation and phenotypes (CoNSeP)~\cite{hover} and multi-organ nucleus segmentation (MoNuSeg)~\cite{monuseg} datasets are used. CoNSeP is proposed in~\cite{hover} and comprises 41 H\&E histological image tiles (1000$\times$1000 pixels) from colorectal adenocarcinoma (CRA) WSIs, where 27 images used for training and 14 images used for testing. MoNuSeg consists of 30 images (1000$\times$1000 pixels) and the images are split in the same way as~\cite{hover} (16 for training and 14 for testing).

To evaluate the impact on higher order task, CGC-Net~\cite{cgc} is used as a tissue classification model. However, in~\cite{cgc}, the dataset used for experiments is not publicly available. To address this problem, another dataset, Camelyon16~\cite{camelyon}, is used. Camelyon16 contains hundreds of H\&E lymph node WSIs with tumor area annotations. 30 WSIs for each class (tumor and normal) are selected from it, and patches (1792$\times$1792 pixels) are extracted from the WSIs. For tumor cases, only patches containing more than 30\% tumor area in the center region (896$\times$896 pixels) are extracted. After acquiring tumor patches, total number of patches is divided by 30, and the quotient is set as the number of patches to extract from normal case WSIs. As a result, even number of patches for both classes are extracted and used for classification model training and testing.

\subsection{Trade-off Between Manual Effort and Performance}
The core of this paper is to reduce manual effort for annotation substantially. 
However, it is important to evaluate what is the degradation that can be expected from a nuclei segmentation algorithm for a given level of reduction of annotation effort.
To explore the trade-off between manual effort and performance, 55 models with different ratios of mixed labels (true labels and pseudo-labels) are trained on MoNuSeg. 
Under each given ratio, the defined percent of patches are randomly chosen from training set and their true labels are replaced with pseudo-labels. 
For each ratio, 5 models are trained and pseudo-label patches are re-sampled for each model. Performance of each ratio is characterized by mean and standard deviation of 5 models.

Figure~\ref{fig2} shows the curve of two metrics (DICE and $\mathrm{DQ}_{point}$) of the 11 ratio settings. Although the resulted curves are almost linear, it is worth noticing that training with about 70\% labels as pseudo-labels yields only about 6\% degradation in DICE and $\mathrm{DQ}_{point}$ metrics. This implies that by randomly choosing 30\% of the samples and annotating them with masks, it can achieve a relatively acceptable performance drop while saving 70\% of exhaustive labeling time, which is disproportionately more time-consuming than point annotation.
For the $\mathrm{DQ}_{point}$ metric, the value of adding the dot branch for pseudo label refinement is more apparent when the ratio of pseudo labels is higher. 

\begin{figure}[htb]
\includegraphics[width=\textwidth]{DICE-DQpoint-eps-converted-to.pdf}
\caption{(a) DICE and (b) $\mathrm{DQ}_{point}$ scores of models trained with different ratio of pseudo-labels and true labels. The x-axis shows the percentage of pseudo-labels, PseudoRate = 0 is equivalent to HoVer-Net, while PseudoRate = 1 is equivalent to the modified approach, NINEPINS. Red error bars are calculated with 5 models in each ratio of pseudo-label setting.} \label{fig2}
\end{figure}

\subsection{Robustness Against Point Label Inaccuracy}
\begin{figure}
\centering
\includegraphics[width=0.2\textwidth]{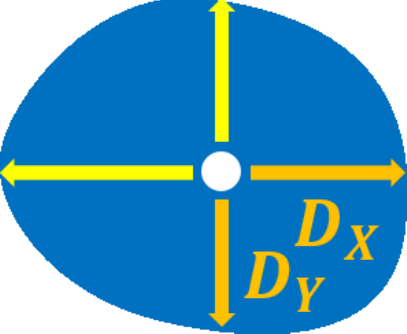}
\caption{Distance to boundaries of the nucleus. $D_X$ and $D_Y$ are, respectively, the minimum horizontal and vertical distances to the nucleus centroid.} \label{fig6}
\end{figure}

NINEPINS introduces a series of modifications to HoVer-Net in order to train using point annotations only. 
Therefore it is necessary to evaluate the robustness of NINEPINS when the point annotations are shifted from their original location. 
In that regard, the point annotations are randomly shifted while keeping the point within the labeled instance, according to figure \ref{fig6}.  
Starting from the centroid of the nucleus segmentation, $D_X$ and $D_Y$ are chosen as the maximum perturbation distance and a parameter $\epsilon$ is introduced as the perturbation strength. 
Random numbers $P_X$ and $P_Y$ are drawn from Gaussian distributions to be used as displacements (See equations \ref{eq:shift_x} and \ref{eq:shift_y}). 
The point shifted by ($P_X$, $P_Y$) is checked for whether it is still inside the nuclear region, and displacements will be re-drawn if it is not.
The shifting method is run 5 times for each perturbation strength to create different point labels.
Afterwards, these point labels are used to generate pseudo-labels and further train HoVer-Net. The dataset used in the experiments is CoNSeP.

\noindent
\begin{equation}
P_X \sim \mathcal{N}(0,\,(\epsilon D_X/3)^{2})\,.
\label{eq:shift_x}
\end{equation}
\begin{equation}
P_Y \sim \mathcal{N}(0,\,(\epsilon D_Y/3)^{2})\,.
\label{eq:shift_y}
\end{equation}

\begin{figure}[htb]
\centering
\includegraphics[width=\textwidth]{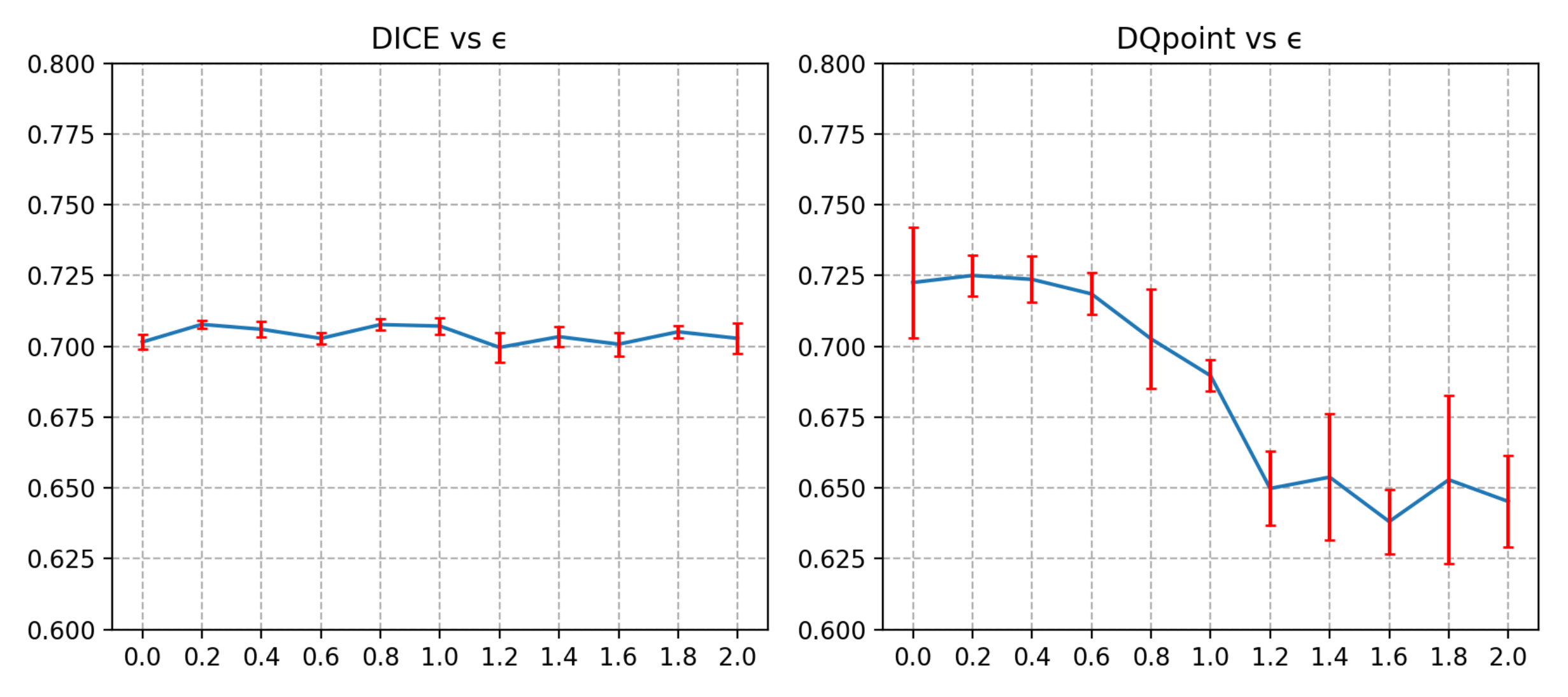}
\caption{(a) DICE (b) $\mathrm{DQ}_{point}$ scores of model trained with point labels shifted with different $\epsilon$. Red error bars are calculated with 5 models in each $\epsilon$ setting.} \label{fig7}
\end{figure}

According to figure~\ref{fig7}~(a), the segmentation ability of our method barely changes when $\epsilon$ value changes. Even when $\epsilon$ value is increased beyond reasonable range ($\epsilon$ $\leq$ 1), the DICE score stays stable. In contrast, figure~\ref{fig7}~(b) shows that the $\mathrm{DQ}_{point}$ score is influenced by strong shifting. The reason might be that the distance map labels depended highly on the point positions and thus affect the detection ability. However, within a reasonable range ($\epsilon$ $\leq$ 1), the $\mathrm{DQ}_{point}$ score only varies in a small range (about 0.03). As a conclusion, the performance of proposed method is robust against inaccurate point labels when the points are still around nuclear centroids, and it degrades with a lower bound when most of the points are far from the centroids and probably located on nuclear boundaries.

\subsection{Impact on Higher Order Task}
\begin{figure}[hbt]
\centering
\includegraphics[width=\textwidth]{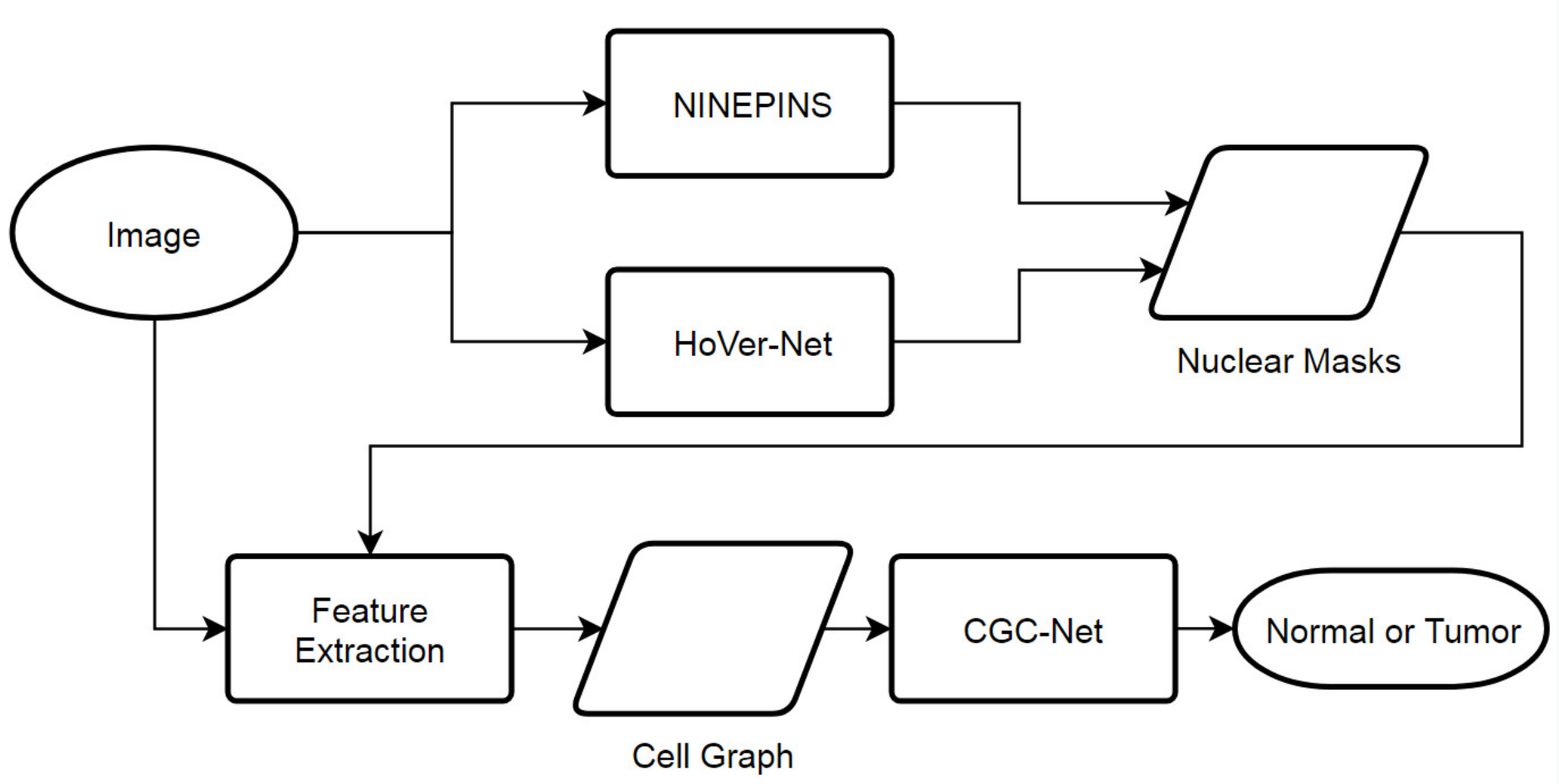}
\caption{Tissue classification pipeline.} \label{flow}
\end{figure}

To compare the performance of classification task using HoVer-Net as instance segmentation model with the one using NINEPINS, we evaluate the results from training CGC-Net for classification. Since the input of CGC-Net are cell graphs which are generated from the instance nucleus masks and the original images, the instance segmentation model needs to have good generalization ability so that it is trained with some other dataset but can be used to do inference on this classfication dataset. 
In our experiments, both instance segmentation models are trained with MoNuSeg dataset and used to produce nucleus mask for the patches extracted from Camelyon dataset. Subsequently, the cell graphs constructed from instance masks and images are used for CGC-Net training. Figure~\ref{flow} shows the flow chart of the pipeline. Similar to the approach in~\cite{cgc}, 3-fold cross-validation is used in the experiments and the patches in each corresponding fold are the same across cases using HoVer-Net and NINEPINS. Different from~\cite{cgc}, Camelyon has only two classes, normal and tumor, instead of three classes (normal, low grade, high grade). Thus the final fully-connected layer of CGC-Net is modified to have two nodes. Table~\ref{tab1} shows the results of  validation accuracy of individual fold and the average over them. 
The results show that a segmentation degradation with point annotations of up to 10\% in terms of DICE or $\mathrm{DQ}_{point}$ metrics does not always imply the same level of performance degradation in higher order tasks.

\begin{table}
\centering
\caption{Classification accuracy of each fold using HoVer-Net and NINEPINS.}\label{tab1}
\begin{tabular}{|c|c|c|}
\hline
Fold  & HoVer-Net & NINEPINS\\
\hline
1 & 88.15\% & 94.20\%\\
2 & 96.00\% & 90.60\%\\
3 & 96.39\% & 90.04\%\\
\hline
$\mu \pm \sigma $ & 93.51 $\pm$ 3.90 \% & 91.60 $\pm$ 1.85 \%\\
\hline
\end{tabular}
\end{table}

\section{Conclusion}

In this paper we have proposed a modification of the HoVer-Net architecture for nuclei instance segmentation to enable traininig using only point-based annotations. 
To compare with HoVer-Net, the performance was evaluated using the same set of pseudo labels for both HoVer-Net and our modification. In these cases, there was no noticeable difference between the performance of the original HoVer-Net and the modified version. However, results showed a slight improved performance for our  method when the ratio of pesudo labels was higher. 
The modifications introduced make our method not only suitable for point annotations, but also robust against shifts in these annotations. 
Finally, upon evaluation of the impact on a higher order task, results indicate that performance only slightly degrades when using much simpler method to annotate nuclei. 
We take this as an indication that weakly supervised or unsupervised methods, if carefully designed, should allow experts to annotate more complex tasks and leave the simpler ones to automated methods.

\bibliography{mybibliography}
%






\end{document}